\documentclass[12pt]{article}

\usepackage{epsf}

\usepackage{graphicx}%

\begin{document}

\begin{center}
{\bfseries A Scientific Analysis of the Preprint arXiv:1301.1828v1  [nucl-th]  
   }

\vskip 5mm

K.A. Bugaev$^{1 \dag}$, D.R. Oliinychenko$^{2}$  and   A.S. Sorin$^{2}$

\vskip 5mm

{\small
(1) {\it
Bogolyubov Institute for Theoretical Physics, Metrologichna str. 14$^B$, Kiev 03680, Ukraine
}
\\
(2) {\it
Bogoliubov Laboratory of Theoretical Physics, JINR, Joliot-Curie str.  6,   Dubna, Russia
}
\\
$\dag$ {\it
E-mail: Bugaev@th.physik.uni-frankfurt.de
}}
\end{center}

\vskip 5mm

\begin{abstract}
Below we analyze  the `critic' statements made in the Preprint arXiv:1301.1828v1  [nucl-th].
The doubtful scientific argumentation  of  the authors of the Preprint arXiv:1301.1828v1  [nucl-th]  is  also discussed.
\end{abstract}

\vskip 3mm

{\large  \bf 1. Introduction.} 
Recently  there appeared  the Comment   arXiv:1301.1828v1  [nucl-th]  A. Tawfik, E. Gamal and H. Magdy 
\cite{Tawfik:13} to our  recent work \cite{KABOliinychenko:12}. Since   this Comment   is   based on an obsolete and highly unrealistic version of the hadron resonance gas model (HRGM),   then   the critical remarks  presented  in \cite{Tawfik:13}  look  like  an attempt to `prove'  that  the results of  more elaborate 
and more realistic versions of  the HRGM \cite{KABOliinychenko:12,KABAndronic:05, KABAndronic:09}
are wrong.
This very fact  forced  us  to   analyze   the main statements of  the opus \cite{Tawfik:13} in order to clearly demonstrate  its original   pitfalls. \\

The main `critique'  statements made in the Comment  \cite{Tawfik:13}  are as follows:
\begin{description}
\item[No 1.] The authors of  \cite{KABOliinychenko:12} ``entirely disregarded the experimental results in baryo-chemical potentials $\mu_b$ and their corresponding temperatures $T$".  

\item[No 2.] The chemical freeze out criterion of constant  entropy  per hadron $\frac{s}{\rho_p} \simeq 7.18$ which   was found to be robust  in \cite{KABOliinychenko:12}  is simply wrong.

\item[No 3.]  A few popular chemical freeze-out criteria (see later)  agree well with the condition  $s/T^3 = 7$
suggested in \cite{Tawfik:06a, Tawfik:06b}.

\item[No 4.]  In addition the authors of  the Comment \cite{Tawfik:13} claim that a criterion of constant  entropy per hadron is an ad hoc one and it has no explanation.  

\end{description}

All other statements made in the Comment  \cite{Tawfik:13} are hard to discuss since the 
above statements No 1-4 clearly demonstrate us that  the authors of the Comment \cite{Tawfik:13} do not know  about the recent development of the  HRGM  made in \cite{KABOliinychenko:12,KABAndronic:05, KABAndronic:09}.
Hence,  we  concentrate only  on  the  statements No 1-4 listed above.\\

{\large  \bf 2. Scientific vs. nonscientific statements in \cite{Tawfik:13}.}
First of all it is necessary to remind that, in contrast to the statement {\bf No 1} of the authors of the Comment \cite{Tawfik:13}, there are NO any ``experimental results in baryo-chemical potentials $\mu_b$ and their corresponding temperatures T" at chemical freeze out or   at any other  stage of heavy ion reaction. This is because such quantities 
(and all other thermodynamic quantities) cannot be directly measured in the experiments. All of them require some model, which, with some success, may allow us to extract the particle or charge densities, or  $\mu_b$ and  $T$ by fitting the experimental data  on hadron multiplicities by a model. If a model has a realistic physical input,  then an extracted information is a reliable one, otherwise any result can be obtained.  Therefore,  the statement {\bf No 1} is a nonscientific one. 

Furthermore, due to the absence of  the first principle theoretical arguments in the phenomenological analysis of the experimental data any statement like  {\bf  No 2}  that some phenomenological result is wrong indicates that the authors of the Comment \cite{Tawfik:13} (HRGM1 hereafter)  have NO any solid scientific arguments against the results of work \cite{KABOliinychenko:12} (HRGM2 hereafter).  
A detailed analysis of their model outlined in \cite{Tawfik:06a} completely supports such a conclusion. 
The worst, however, is that the authors of the Comment \cite{Tawfik:13} claim wrong  not only the results of  
\cite{KABOliinychenko:12}, but  many years of research  to formulate the most successful version of the HRGM  \cite{KABAndronic:05,KABAndronic:09} (HRGM3 hereafter)  on which our formulation HRGM2  \cite{KABOliinychenko:12} is mainly  based. Although the  particle table and the treatment of  the resonance width in  the HRGM2 \cite{KABOliinychenko:12} are  slightly different   compared to the HRGM3 \cite{KABAndronic:05,KABAndronic:09} the main results of these models  are very close to each other.

Usually, the HRGM  is used to extract the thermodynamic quantities from the hadron yields measured under certain conditions (at midrapidity or in $4 \pi$ solid angle). At present there are many different formulations of  the HRGM, but the most successful one, the HRGM3,  was developed by  A. Andronic, P. Braun-Munzinger and J. Stachel in \cite{KABAndronic:05,KABAndronic:09}. A great success of  the HRGM3 \cite{KABAndronic:05,KABAndronic:09}  is naturally explained by its realistic features. The most important of them are as follows:

{\bf I. The presence of the hard core repulsion between hadrons.}   This feature is of a principal importance \cite{Kapusta:81,Cleymans:06}, since in the absence the hard core repulsion between hadrons the hadronic pressure becomes so huge that there is no transition to the quark gluon matter, if all hadrons with masses up to 2 GeV are accounted. Evidently, such a model simply contradicts to QCD and, hence, it cannot be used at temperatures exceeding the pion mass. The last statement is based on  the fact that the hard core repulsion essentially reduces the particle densities compared to the ideal gas.  See, for instance,  Fig. 3 in \cite{Begun:12},  where it is shown that such a  reduction can be up 90 \%  (!) and hence an  ignorance of the  hadron hard core repulsion may lead to unrealistic values of such thermodynamic parameters as chemical freeze out volume or ratios between the yields of the most abundant hadrons (pions) and the less abundant  ones (multistrange baryons). 

{\bf II.  All hadronic resonances with masses up to 2.5 GeV should be accounted.}  This is necessary to successfully describe the hadronic multiplicities for the center of mass energies per nucleon $\sqrt{s_{NN}} > 6$ GeV \cite{KABAndronic:05,KABAndronic:09}.  
It is also evident that  the Properties I  and II are closely related, because, if  more resonances are 
taken into  account,  then  the stronger deviation from the mixture of  ideal gases  should be expected.  

{\bf III. It is also important that wide hadronic resonances are accounted  in a proper way.}  In other words, the wide resonances should not be treated as stable particles, but  their spectral functions up to a  threshold of  the leading  channel of decay should be implemented into a model. Usually,  it is believed that the width of wide resonances is important at low temperatures \cite{KABAndronic:05} only.  However, recently \cite{Bugaev:12a,Bugaev:12b} it was shown that the heavy and wide resonances should be taken into account  up to temperatures of about 170 MeV.

{\bf IV.  The full hadronic multiplicities at chemical freeze out should take into  account  both the thermal hadronic yields  and the yields coming from the decays of heavier resonances.}  Otherwise it is impossible to describe the measured  hadronic multiplicities.  For instance, it is well known that without inclusion of  $\sigma(600)$ meson into the  HRGM it is hard to correctly describe the pion yield at energies  $\sqrt{s} < 6 $ GeV  \cite{KABAndronic:05} because just this meson alone provides up to 5 \% of total pion yield at these energies.

{\bf V. The conservation laws.}  Usually only the strangeness conservation is taken into account explicitly by  finding out the chemical potential of the strange charge from a condition of vanishing strangeness. 

As one can judge from \cite{KABOliinychenko:12}, one of the main purposes of this paper was to demonstrate that a form of 
conservation laws (5) and (6) suggested in \cite{KABAndronic:05} and used afterwards leads to unrealistically small volumes
at chemical freeze out (see Fig. 3 in \cite{KABOliinychenko:12}). The critique is strong, but convincing. Moreover, as one can see from \cite{KABAndronic:12}
the critique put forward in \cite{KABOliinychenko:12} is accepted and the corresponding conservation laws are modified. \\

\begin{figure}[ht]
\centerline{\includegraphics[width=7.7 cm]{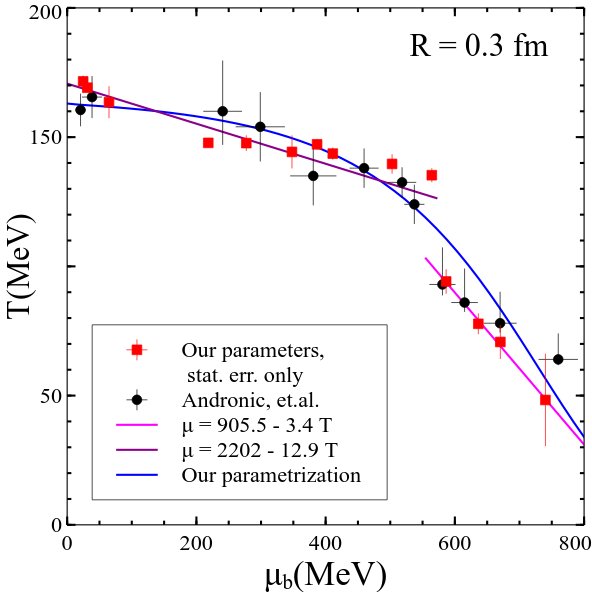}   
\includegraphics[width=7.7 cm]{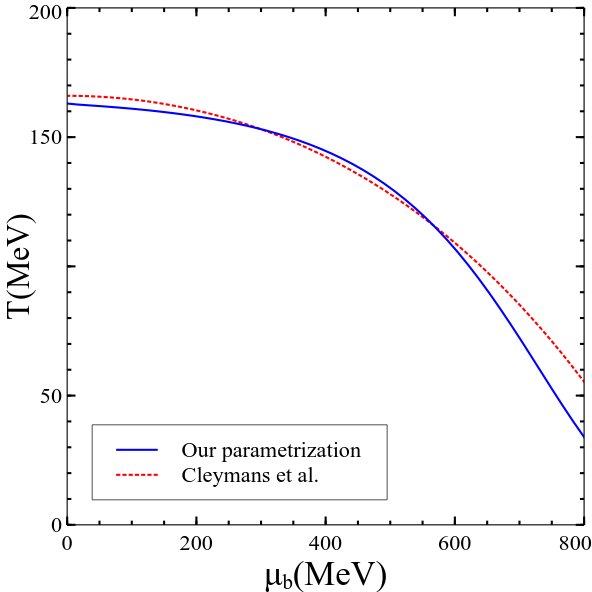} 
}
 \caption{Chemical freeze out temperature dependence on  the baryonic chemical potential $\mu_b$.
 The symbols in the left panel correspond to the fit of hadron yield ratios obtained in \cite{KABOliinychenko:12}  (squares) and in \cite{KABAndronic:05} (circles) for the same value of the hard core radius of all hadrons $R=0.3$ fm. The  solid curve in the left panel  is  
 a  fit  to the results   of  \cite{KABOliinychenko:12} and  \cite{KABAndronic:05} by Eq.   (\ref{EqII}).
The dashed and solid curves in the right panel  correspond to 
 the equations  (\ref{EqI}) and  (\ref{EqII}), respectively.
 In fact, the straight lines with the parameters specified in the left panel, describe well the $\mu_b$ dependence of the chemical freeze out  temperature.   
   }
 \label{fig1}
\end{figure}

{\large  \bf 3. The doubtful scientific  argumentation  in  \cite{Tawfik:13}.}
The  HRGM1 used by the authors of the Comment \cite{Tawfik:13} is highly unrealistic since it does not possess  the Properties I-IV and, hence, any physical conclusion drawn out of  it is simply unrealistic.  Moreover, the main critique of the authors of the Comment \cite{Tawfik:13} is based on the parameterization \cite{Cleymans:06}
\begin{equation}\label{EqI}
T (\mu_b) = a - b\, \mu_b^2 - c \, \mu_b^4 \, ,
\end{equation}
with 
$a = 0.166 \pm 0.002$ GeV, $ b = 0.139 \pm 0.016$ GeV$^{-1}$ and $c = 0.053 \pm 0.021$ GeV$^{-3}$.
The parameterization (\ref{EqI}) is based on a compilation of results of a few models and not all of them  are 
supplemented  by   the Properties I-IV.
As it is clearly seen from Fig. \ref{fig1}  the chemical freeze out temperature dependence of the models \cite{KABOliinychenko:12, KABAndronic:05}  differs from  (\ref{EqI})  and hence any  critique  of the Comment \cite{Tawfik:13} based on the equation (\ref{EqI})  is  not eligible.  

A functional dependence relating  the values of  $T$ and $\mu_b$ at chemical freeze out which approximately describes the results
found by the HRGM2 \cite{KABOliinychenko:12} and the HRGM3  \cite{KABAndronic:05} is as follows 
\begin{equation}\label{EqII}
T (\mu_b) = \frac{T_0}{1 +  \frac{ 2.44}{\left[ \ln\left(  \frac{1}{a} \left[  \frac{\mu_0}{\mu_b} - 1  \right] \right)   \right]^4} } \, , \quad {\rm for} \quad \mu_b \le  750 ~\rm MeV\, ,
\end{equation}
where $T_0 \simeq 163$ MeV, $a \simeq 0.31$ and $\mu_0 \simeq 1407$ MeV.  At first glance  it seems that  the curves 
defined by the equations  (\ref{EqI}) and  (\ref{EqII}) and shown in Fig. 1   do not differ essentially. Indeed, for $\mu_b < 750$ MeV
the difference of two freeze out temperatures is below 25 MeV. However, due to the absence of Properties I-IV 
in the the HRGM1  employed by the authors of  \cite{Tawfik:13} the corresponding  particle densities found within   the HRGM1  \cite{Tawfik:13} and   the HRGM2 \cite{KABOliinychenko:12} may differ essentially.  A more accurate parameterization  $\mu_b (T)$ at  the chemical freeze out found in  \cite{KABOliinychenko:12}  is given in the left panel of  Fig.  \ref{fig1}.

In `criticizing' the chemical freeze out condition $\frac{s}{\rho_p} \simeq 7.18$ \cite{KABOliinychenko:12} 
the authors of  \cite{Tawfik:13} use the   doubtful scientific  argumentation. First of all, they simply 
ignore the results of  the lower panel of Fig. 6 in  \cite{KABOliinychenko:12} which clearly demonstrates that such a criterion is valid even 
at low center of mass energies of collision $\sqrt{s_{NN}} \ge 2.3$ GeV, i.e. at large values of baryonic  chemical potential  
$\mu_b > 500$ MeV.  Instead, the authors of the Comment  \cite{Tawfik:13} claim that ``It is obvious that 
$s/n$ never reaches 7.18 at $\mu_b > 500$ MeV'', forgetting to mention  that this conclusion is obtained not within the realistic HRGM2   \cite{KABOliinychenko:12}, but within the unrealistic HRGM1   \cite{Tawfik:13,Tawfik:06a, Tawfik:06b}.  

The second example of  the doubtful scientific  argumentation used by the authors of the Comment  \cite{Tawfik:13} is  as follows. In order  to  prove the validity of the statement {\bf No 2} the  authors of the Comment  \cite{Tawfik:13}  substitute  the particle number density $\rho_p$ by the baryonic charge density $n$. In fact, Sect. III of the Comment \cite{Tawfik:13} is called as `PHYSICS OF CONSTANT ENTROPY PER NUMBER DENSITY', but as one can judge from the equation (3) in \cite{Tawfik:13} which is  written as 
\begin{equation}\label{EqIII}
\frac{s}{n} = \frac{1}{T} \left( \frac{p}{n} + \frac{\epsilon}{n} - \mu_b    \right) \,,
\end{equation}
either $n$   is a baryonic charge density and, hence,  the authors of the Comment are  criticizing not the 
condition of constant entropy per particle, or  $n$ is, indeed, the particle number density, but then the equation 
(3) in the Comment \cite{Tawfik:13} has nothing to do with the standard  thermodynamics. Since the  authors
of  \cite{Tawfik:13} failed to specify their notations used in (3),  here  we also assumed that they
consider $p$ as the system pressure and $\epsilon$ as its energy density.

The third example of  the doubtful scientific  argumentation  in  the Comment   \cite{Tawfik:13}
requires a special attention. In order to `prove'  the validity of their statement {\bf No 2}
the  authors of  the Comment   \cite{Tawfik:13} simply   extrapolate (with the help of  the parameterization (\ref{EqI}) !)  the HRGM1  results of  \cite{Tawfik:13, Tawfik:06a, Tawfik:06b}  to the chemical freeze out temperatures somewhat  well  below  50 MeV  and demonstrate  that the entropy per baryonic charge is essentially larger that 7.18.
From such a procedure the authors of the Comment  \cite{Tawfik:13} conclude that the chemical freeze out criterion $\frac{s}{\rho_p} \simeq 7.18$  \cite{KABOliinychenko:12}  cannot be used at AGS and SIS energies.   However, we  have to stress  here  that to our best knowledge none of the realistic  thermal models, including the HRGM3,  which are able to describe the particle ratios at SIS energies  $\sqrt{s_{NN}} = 2. 24 $  GeV  and $\sqrt{s_{NN}} = 2. 32 $  GeV ever showed the chemical freeze out temperatures below $49$ MeV (see, for instance, 
\cite{Cleymans:06}). Hence, there is no need to worry about the behavior of the ratio $\frac{s}{\rho_p} $ at $T < 50$ MeV! \\

{\large  \bf 4. A special role of the chemical  freeze out criterion $s/T^3 = 7$.} The authors of the Comment \cite{Tawfik:13} 
are considering a few traditional chemical freeze out criteria, namely, of constant energy per particle, but  they write it as  $\epsilon/ n \simeq 1$ GeV,  and of constant number of baryons and antibaryons $n_b + n_{\bar b} \simeq 0.12$
fm$^{-3}$, but the main attention is paid to the criterion $s/T^3 = 7$ suggested in \cite{Tawfik:06a,Tawfik:06b}. It is necessary to remind that the  criterion  $s/T^3 = 7$ was heavily  criticized already in  \cite{Cleymans:06}, where it was demonstrated that an inclusion of the hard core repulsion into  the HRGM1  essentially modifies the relation (\ref{EqI})  between the chemical freeze out  parameters for the criterion $s/T^3 = 7$. This is clearly seen in Fig 2. 

\begin{figure}[ht]
\centerline{\includegraphics[width=8.8 cm]{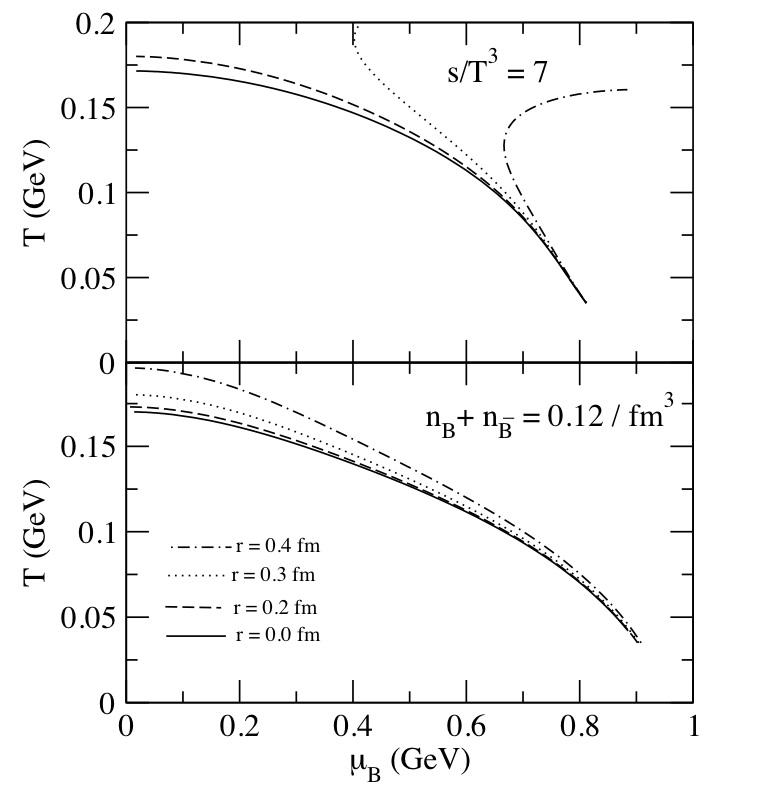}   
}
 \caption{The effect of excluded volume corrections on the constant $n_B + n_{\bar B}$ (bottom) 
 and constant $s/T^3$  
(top) freeze-out criterion. This figure  is taken from the preprint 
J. Cleymans, H. Oeschler, K. Redlich, and S. Wheaton,
arXive:hep-ph/0511094v2 of Ref.  \cite{Cleymans:06} in order to demonstrate the unrealistic 
behavior of the chemical  freeze-out criterion   $s/T^3 = 7$, if the hard core repulsion is taken into account. For the hard core radius $R=0.3$ fm which was used in \cite{KABOliinychenko:12,KABAndronic:05,KABAndronic:09} to fit the data  the chemical  freeze-out criterion   $s/T^3 = 7$ does not work. 
   }
 \label{fig2}
\end{figure}

Probably, the authors of the Comment  \cite{Tawfik:13}  think it is a great advantage of their model that all the chemical freeze out criteria shown in the left panel of Fig. 2  calculated in  \cite{Tawfik:13} at the curve $s/T^3 = 7$  demonstrate a constant behavior for all values of the baryonic chemical potential from $\mu_b \simeq 5$ MeV to  $\mu_b \simeq 10000$ MeV. 
We, however,  would like to remind  that at so huge values of the baryonic chemical potential ($\mu_b \gg 1000$ MeV) there is no 
reason to discuss both the chemical freeze out and the HRGM, since according to the contemporary QCD 
 there should exist other state of matter at this region and, hence, the hadron resonance gas is simply inapplicable. 

Also it is necessary to stress   that the chemical freeze out criterion  $s/T^3 = 7$ is not  observed 
in the HRGM2  \cite{KABOliinychenko:12} and in the HRGM3 \cite{KABAndronic:05} and a similar conclusion  is also confirmed  by the recent analysis of  \cite{Begun:12}.   
In Ref. \cite{Begun:12} the parameterization (1) is used for the HRGM which is similar to the HRGM1  \cite{Tawfik:13}.
As one can see from the right panel of  Fig. 2 in  \cite{Begun:12} in this case 
 the ideal gas model gives  $s/T^3 \simeq  7$ for the lab energies of collision above 4 GeV per nucleon and just for a strangeness suppression factor equal to 1 (no suppression), while   for smaller lab energies the ratio $s/T^3$  is essentially  larger than 7! 
 If, however, one introduces the strangeness suppression factor dependence as suggested in \cite{Becattini:06}, then $s/T^3 = 6$ for all lab energies above 8 GeV per nucleon. Finally, if one employes  the parameterization (1) for the HRGM with the hard core repulsion, then, as one can see from the right panel of Fig. 4 in  \cite{Begun:12},   
 $s/T^3$ varies from 3.6 to 6, depending on the set of hard core radii.
 
 Therefore, in order to prove the claim {\bf No 3} the authors of the Comment \cite{Tawfik:13} forget about the parameterization (1) which they  used to `criticize' the   HRGM2 and  HRGM3 results   presented in  \cite{KABOliinychenko:12}. Thus,  the authors of the Comment  \cite{Tawfik:13} use the double standards.
 
Finally, before claiming that a criterion of constant entropy per hadron is an ad hoc one (claim {\bf No 4})   it would be nice, if  the authors of the Comment \cite{Tawfik:13} could   follow their own advice in the first place and could not  not  ignore  the existing literature on this subject.  Probably,  the authors of the Comment \cite{Tawfik:13} should have looked into a recent work \cite{Bugaev:12b} to study the suggested explanation for a criterion of constant  entropy per hadron. \\

{\large \bf 5. Conclusions.} The above analysis clearly shows us that  the Comment \cite{Tawfik:13} 
 lacks  any new result  and its authors  are  trying to prove an  impossible,
 namely that their obsolete formulation of the HRGM1 has some advantages over 
 more elaborate ones. 
 In contrast to their own  calls to lift up the scientific standards, 
 the authors of the Comment  \cite{Tawfik:13} 
 use  the doubtful scientific argumentation
 to `prove' the validity of unrealistic model of Refs. 
 \cite{Tawfik:13,Tawfik:06a,Tawfik:06b} and 
 to claim wrong  the results of  the advanced  HRGM formulations worked out in  \cite{KABOliinychenko:12,KABAndronic:05,KABAndronic:09}.


\end{document}